\documentclass[aps,twocolumn,showpacs,pra,amssymb, amsmath]{revtex4}
\usepackage{amsmath,latexsym,amssymb}
\usepackage{graphicx}
\usepackage{amsmath}
\usepackage{latexsym}
\usepackage{amssymb}
\usepackage{bm}

\begin{document}

\title{Nonlocality improves Deutsch algorithm}

\author{Koji Nagata}
\affiliation{ Department of Physics, Korea Advanced Institute of
Science and Technology, Daejeon 305-701, Korea}
\author{Sangkyung Lee}
\affiliation{ Department of Physics, Korea Advanced Institute of
Science and Technology, Daejeon 305-701, Korea}
\author{Jaewook Ahn}
\affiliation{ Department of Physics, Korea Advanced Institute of
Science and Technology, Daejeon 305-701, Korea}

\date{\today}

\begin{abstract}
Recently, [{arXiv:0810.3134}] is accepted and published.
We show that the Bell inequalities lead to a new type of
linear-optical Deutsch algorithms. We have considered a use of
entangled photon pairs to determine simultaneously and
probabilistically two unknown functions. The usual Deutsch algorithm
determines one unknown function and exhibits a two to one speed up
in a certain computation on a quantum computer rather than on a
classical computer. We found that the violation of Bell locality in
the Hilbert space formalism of quantum theory predicts that the
proposed {\it probabilistic} Deutsch algorithm for computing two
unknown functions exhibits at least a $2\sqrt{2}(\simeq 2.83)$ to
one speed up.

\end{abstract}

\maketitle

\section{Introduction}
Recently, \cite{NagataNakamura} is accepted and published.
Quantum information processing and quantum computing have attracted
much interest in science community because of their novel usage of
quantum mechanics in technological applications. %
Many ideas of quantum processors and computers were experimented in
many architectures of physical systems, including ion traps, neutral
atoms, nuclear spins in magnetic resonance, semiconductor quantum
dots, and super-conducting
resonators~\cite{IONTRAPQC,ATOMQC,NMRQC,QuantumDotQC,SCQC}. Still
the progress has been limited to a few qubit operations and the
performance is far from being practical. The ability of quantum
computers in outperforming their classical counterparts has not been
demonstrated.

In many physical systems, the preparation and change of quantum
states, or how to prepare entanglements and how to maintain
coherence, is a more difficult task than how to wire logics
quantum-mechanically. However, in quantum computing with linear
optical systems, it is relatively easy to deal with entanglement and
decoherence. For this reason, linear optical quantum computings or
the linear interactions of photons with matters are often adopted
for the implementation of $N$-qubit quantum
algorithms~\cite{AhnScience2000, BhattacharyaPRL2002}. Especially
entanglement is an important aspect that a quantum mechanical device
can have and the quantum information carried by an entangled state
like Einstein, Podolsky, and Rosen (EPR) state overcomes some of the
limitations of classical information used in communication and
cryptography~\cite{NC,Galindo,CRYPTOGRAPHY,BRUKNER_PRL}. Recently,
there have been several attempts to use single-photon two-qubit
states for quantum computing. Oliveira {\it et al.} implemented the
Deutsch algorithm with polarization and transverse spatial modes of
the electromagnetic field as qubits~\cite{Oliveira}. Single-photon
Bell states were prepared and measured by Kim~\cite{Kim2003}. Also
the decoherence-free implementation of Deutsch algorithm using such
single-photon two logical qubits~\cite{Mohseni2003}. Although such a
single-photon two-qubit implementation is not scalable, the quantum
gates necessary for information processing can be implemented
deterministically using only linear optical elements.

Often a demonstration of quantum algorithm, which is performed with
possibly mixed quantum states, is presented without a proper
mathematical theory for the analysis of experimental data, or with a
rather complicated quantum tomographical state analysis. However, if
the output state of an experiment under study should be an entangled
state, we can choose to use Bell inequalities~\cite{BELL}. It is a
sufficient condition to demonstrate a negation of Bell locality and
the detection of entanglement. We can consider the following
question. Is there a relationship between the negation of Bell
locality and the performance of such quantum algorithm that can be
implemented by a single photon? Interestingly the answer is yes.

In this paper we have devised an experimental scheme to obtain
simultaneous and nonlocal answers from Deutsch problem with two
unknown functions. Especially we elaborate the use of
\emph{entanglement} in processing this quantum algorithm. The
advantage of using an EPR pair of photons (two-photon two-qubit
states) in quantum computing algorithm is analyzed with Bell
inequalities in quantum theory. We show that a set of answers of the
given Deutsch problem, with two unknown functions, statistically
shows a violation of Bell locality in the Hilbert space formalism of
quantum theory. It turns out that the negation of Bell locality
exhibits a $2\sqrt{2}$ to one speed up at least. We, thus, observe a
highly nonlocal effect and it leads the entangled answers in a
network of pair quantum computers to provide enhanced information
compared with its classical counterpart. An important note here is
that there must be probabilistic errors in answers which appear due
to the imperfection of the photon detection and defects in optical
devices. Thus it is necessary to take into account many measurements
and what we can do is only to analyze probabilistically in real
experimental situation. By applying the maximum likelihood
principle, the answer to the Deutsch problem is estimated
probabilistically.

In the following sections, we introduce a method of linear optical
quantum computing of Deutsch problem. In the
section~\ref{entanglement2}, the method that utilizes two-photon
two-qubit entanglement is discussed. This method allows statistical
analysis of the average value of Bell operator. In order to overcome
the imperfection of the photon detection and possible defects in
optical devices, the fidelity \cite{NAGATA1} of the method is
analyzed. During the analysis of Bell operator or the fidelity to
EPR state in Sec.~\ref{witness}, the separable states are
distinguished from the entangled states. It is well known that the
fidelity which is larger than $1/\sqrt{2}$ to EPR state is a
sufficient condition for a negation of Bell locality in the Hilbert
space formalism of quantum theory. We can say that our Deutsch
scheme succeeds with the probability of the value of the lower bound
of fidelity at least under the condition where the fidelity is
larger than $1/\sqrt{2}$. A short summary and conclusion follows in
Sec.~\ref{summary}.

\section{Deutsch algorithm with two unknown functions}
\label{entanglement2}

\begin{widetext}
\begin{figure}
\begin{center}
\includegraphics[width=1.0\textwidth]{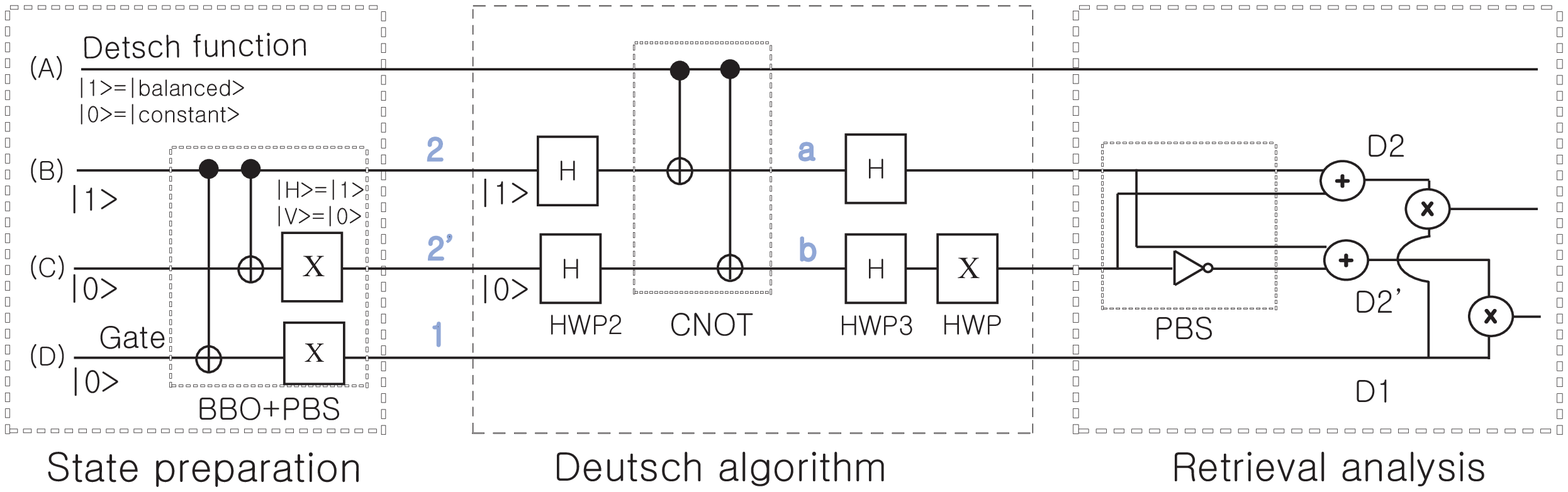}
\caption{The schematic of Deutsch algorithm with a two-photon state.
The classical channel (A) represents the type of the chosen Deutsch
function.  The two quantum channels (B) and (C), in a superposition
single-photon state, are initially prepared in a horizontally
polarized state, $|1\rangle$, and a vertically polarized state,
$|0\rangle$, respectively. The channel (D) is for the extra photon
for a time correlation detection.} \label{schematic}\end{center}
\end{figure}

The Deutsch algorithm determines whether a given function, $f(x)$,
on a binary number, $x$, is either balanced or constant, where the
function is constant if the function outputs for $x=0$ and $1$ are
the same and balanced if not. The unknown function is defined by
\begin{eqnarray}
U_f|x\rangle_{y}\equiv|x\rangle_{y\oplus f(x)}.
\end{eqnarray}
In a classical computer the answer is obtained by calculating $f(0)$
and $f(1)$, while in a quantum computer only a single calculation
with a superposition state of $0$ and $1$ is necessary. Hence, if
$f(H)=f(V)$ then the output label should not depend on $H$ and $V$,
whereas if $f(H)\neq f(V)$ then the output label should depend on
$H$ and $V$.

As shown in Fig.~\ref{schematic}, Deutsch algorithm can be
implemented using a two-photon state. The quantum channels (B) and
(C) represent the horizontally polarized photon, in a superposition
state coherently traveling along two paths 2 and 2', from an EPR
photon pair. The qubits in these channels are to be simultaneously
altered by the control bit in channel (A) that denotes $|1\rangle$
for a balanced function or $|0\rangle$ for a constant function. Then
the final photon state, which is collapsed and detected either at
the detector D2 or D2', reveals the type of the Deutsch function:
whether the unknown function was balanced or constant. The other
photon, in the vertical polarization, is used as a gate function for
the coincidence measurement. This is reminiscent of Oliveira's
scheme in \cite{Oliveira}, except that we discuss quantum-mechanical
advantage of wiring two of those Deutsch algorithms which are
simultaneously processed by an entangled two-photon state.

\begin{figure}
\begin{center}
\includegraphics[width=0.85\textwidth]{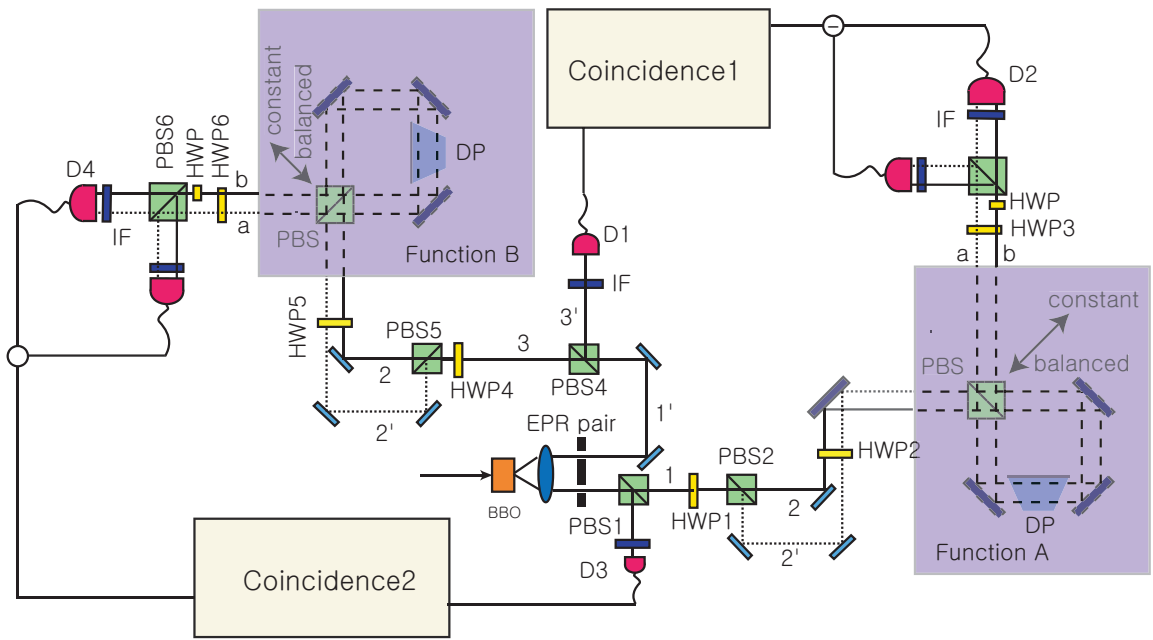}
\caption{The Deutsch algorithm with a two-photon entangled state}
\label{setup3}\end{center}
\end{figure}

\end{widetext}

We describe the experimental scheme shown in Fig.~\ref{setup3}. The
unknown functions $A$ and $B$ in the gray boxes represent the CNOT
gate in Fig~\ref{schematic}. The half-wave-plates labeled from HWP1
to HWP5, oriented at $\theta=22.5^{\circ}$, are the Hadamard gates,
which change a horizontal (vertical) polarization state into a
45$^{\circ}$ (45$^{\circ}$) polarization state. The half-wave-plates
labeled as HWP in Fig.~\ref{setup3} are oriented at
$\theta=45^{\circ}$ and swap the horizontal and vertical
polarizations, like the $X$ gates in Fig~\ref{schematic}. In our
implementations, we assign the truth values 0 and 1 as
$|H\rangle\leftrightarrow |1\rangle,~ |V\rangle \leftrightarrow
|0\rangle$. The initial photon is in the coherent superposition
state of $(|1\rangle_{2}+|0\rangle_{2'})/\sqrt{2}$. Then, the
state evolution after the logic gates $H$ and $V$ depending on the
choice of the unknown Deutsch function, $X$ for the balanced
function or $I$ for the constant function, becomes
\begin{eqnarray}
H \left[ \begin{tabular}{c} $X$ \\ $I$ \end{tabular}
\right]H|1\rangle + \alpha X H \left[\begin{tabular}{c} $X$ \\ $I$
\end{tabular}\right]H|0\rangle=
\left[ \begin{tabular}{c} $1- \alpha$ \\ $1 + \alpha$ \end{tabular} \right]|1\rangle,
\end{eqnarray}
where the detector D2 clicks if $\alpha=1$ and D2' clicks if
$\alpha=-1$.

This scheme is the generalization of the usual Deutsch algorithm in
such a way that we can determine the lower bound of the success
probability of the algorithm and we can see a violation of Bell
locality in the Hilbert space formalism of quantum theory. In this
section, we consider an ideal case, i.e., there is not any
experimental noise to simplify the discussion. However, in real
experimental situations, we have to take error answers into account
due to experimental imperfections. Thus, many runs of experiments
are evidently necessary. Hence, we shall discuss a method using Bell
operators to analyze experimental data. The method of such analysis
will be presented in Sec.~\ref{witness}.

We use Pauli observables for the representation of photon
polarization states as
\begin{eqnarray}
\sigma_z&=&|H\rangle\langle H|-|V\rangle\langle V|, \nonumber \\
\sigma_x&=&|H\rangle\langle V|+|V\rangle\langle H|.
\end{eqnarray}
The initial state of photons is an EPR bi-photon state as
\begin{eqnarray}
\frac{1}{\sqrt{2}}(|V\rangle_1 |H\rangle_{1'}
-|H\rangle_1|V\rangle_{1'}).\label{initial}
\end{eqnarray}
Now we follow the time evolution of each of photons.

\subsection{Deutsch algorithm with a function A}

Assume that the case where the state $|H\rangle_1|V\rangle_{1'}$ is
contributed to our Deutsch algorithm. In this case, the unknown
function (A) is determined as constant or balanced. After HWP1, the
each state of photons becomes
\begin{eqnarray}
|H\rangle_{1}\rightarrow\frac{|H\rangle_{1}+|V\rangle_{1}}{\sqrt{2}},
\quad |V\rangle_{1'}\rightarrow |V\rangle_{D1}.
\end{eqnarray}
Hence the state $|H\rangle_1|V\rangle_{1'}$ becomes
\begin{eqnarray}
\frac{1}{\sqrt{2}}(|H\rangle_1 +|V\rangle_{1})|V\rangle_{D1}.
\end{eqnarray}

A vertically polarized photon is detected by the $D1$ detector.
That is,
\begin{equation}\sigma^{D1}_z=-1.  \label{Pauliz}
\end{equation}
We see that the state of the photons $\frac{1}{\sqrt{2}}(|H\rangle_1
+|V\rangle_{1})|V\rangle_{D1}$ is projected into the following $
 (|H\rangle_1+|V\rangle_{1})/\sqrt{2}$.
 In order to simplify the discussion, in what follows, we consider
this state. The polarizing beam splitter (PBS2) changes each of
states as $ |H\rangle_1 \rightarrow
|H\rangle_2,~|V\rangle_{1}\rightarrow|V\rangle_{2'}$. Hence we have
\begin{eqnarray}
\frac{1}{\sqrt{2}}
(|H\rangle_2+|V\rangle_{2'}).\label{select}
\end{eqnarray}
After the half-wave plate (HWP2) each of the states becomes as
\begin{eqnarray}
|H\rangle_2\rightarrow\frac{|H\rangle_2+|V\rangle_2}{\sqrt{2}},\quad
|V\rangle_{2'}\rightarrow\frac{|H\rangle_{2'}-|V\rangle_{2'}}{\sqrt{2}}.
\end{eqnarray}
From state (\ref{select}), after HWP2, we have
\begin{eqnarray}
\frac{1}{\sqrt{2}} \bigg((\frac{|H\rangle_2+|V\rangle_2}{\sqrt{2}})
+(\frac{|H\rangle_{2'}-|V\rangle_{2'}}{\sqrt{2}})\bigg). \label{SL}
\end{eqnarray}

\begin{figure}
\begin{center}
\includegraphics[width=0.25\textwidth]{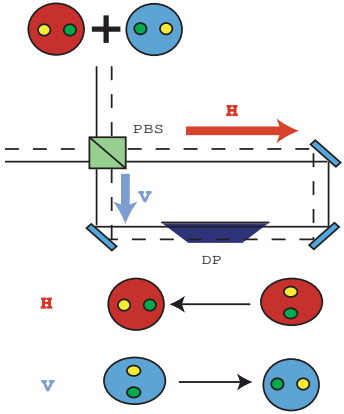}
\end{center}
\caption{An example of polarization-CNOT gates} \label{CNOT}
\end{figure}

A dove prism (DP) is the most important part in implementing the
given functions. The output polarization state of the dove prism
rotates at twice the angular rate of the rotation of the prism
itself. So, if a dove prism inclines at $45^{\circ}$ about a
vertical line, the image rotates at $90^{\circ}$. Now we consider
how a CNOT gate (cf. \cite{CNOT}) is implemented with the space and
polarization degrees of freedom of a photon. As shown in
Fig.~\ref{CNOT}, a horizontally polarized photon is transmitted
through the polarization beam splitter. But a vertically polarized
photon is reflected at the polarization beam splitter. They
propagate along different ways. The angle between a vertical line
and the axis of the dove prism is $45^{\circ}$ in the case of a
horizontally polarized photon, but $-45^{\circ}$ in the other case.
So this configuration implements such changes of photon paths: $2
\rightarrow a$, $2' \rightarrow b$ when the polarization is
horizontal. $2\rightarrow b$, $2' \rightarrow a$ when the
polarization is vertical. Here $a$ and $b$ are the labels for the
paths toward detectors. Output photons labeled by $a$ and $b$ are
detected by $D2$.

%%%%%%%%%%%%%%%%%%%%%%%%%%%%%%%%%%%%
\subsubsection{Balanced function case}
In the balanced function case, the changes of the polarization
states of the photons due to the dove prism are:
\begin{eqnarray}
&&|H\rangle_2\rightarrow |H\rangle_a,~
|H\rangle_{2'}\rightarrow |H\rangle_b,\nonumber\\
&&|V\rangle_2\rightarrow |V\rangle_b,~
|V\rangle_{2'}\rightarrow |V\rangle_a.
\end{eqnarray}
Thus, after the dove prism, the state (\ref{SL}) becomes
\begin{eqnarray}
\frac{1}{\sqrt{2}}
\bigg((\frac{|H\rangle_a+|V\rangle_b}{\sqrt{2}})
+(\frac{|H\rangle_{b}-|V\rangle_{a}}{\sqrt{2}})\bigg).
\end{eqnarray}
After HWP3, we have $(|V\rangle_{a}+|H\rangle_{b})/\sqrt{2}$. After
HWP($\theta=45^{\circ}$), we get
$(|V\rangle_{a}-|V\rangle_{b})/\sqrt{2}$. Hence, we detect only
vertically polarized photons in $D2$. This implies
\begin{eqnarray}
(\sigma^{a}_z=\sigma^{b}_z=)\sigma^{D2}_z=-1.\label{PauliBz}
\end{eqnarray}
Hence from Eqs.~(\ref{Pauliz}) and (\ref{PauliBz})
the value of observable $\sigma^{D1}_z \sigma^{D2}_z$ should be $+1$.
This is one of
the success result of a single run of the experiment, i.e.,
\begin{eqnarray}
\sigma^{D1}_z \sigma^{D2}_z=+1.\label{PaulifinalBz}
\end{eqnarray}
%%%%%%%%%%%%%%%%%%%%%%%%%%%%%%%%%%%%%%%%
\subsubsection{Constant function case}
After the dove prism, polarized photon states changes as follows
\begin{eqnarray}
&&|H\rangle_2\rightarrow |H\rangle_a,~
|H\rangle_{2'}\rightarrow |H\rangle_b,\nonumber\\
&&|V\rangle_2\rightarrow |V\rangle_a,~
|V\rangle_{2'}\rightarrow |V\rangle_b.
\end{eqnarray}
Thus, after the dove prism, the state (\ref{SL}) becomes
\begin{eqnarray}
\frac{1}{\sqrt{2}}
\bigg((\frac{|H\rangle_a+|V\rangle_a}{\sqrt{2}})
+(\frac{|H\rangle_{b}-|V\rangle_{b}}{\sqrt{2}})\bigg).
\end{eqnarray}
After HWP3 we have $ (|H\rangle_{a}+|V\rangle_{b})/\sqrt{2}. $ After
HWP($\theta=45^{\circ}$), we have $
(|H\rangle_{a}+|H\rangle_{b})/\sqrt{2}. $ Hence, we detect only
horizontally  polarized photons in $D2$. This implies
\begin{eqnarray}
(\sigma^{a}_z=\sigma^{b}_z=)\sigma^{D2}_z=+1.\label{PauliCz}
\end{eqnarray}
Hence from Eqs.~(\ref{Pauliz}) and (\ref{PauliCz})
the value of observable $\sigma^{D1}_z\sigma^{D2}_z$ should be $-1$.
This is one of the success result of a single run of the experiment, i.e.,
\begin{eqnarray}
\sigma^{D1}_z\sigma^{D2}_z=-1.\label{PaulifinalCz}
\end{eqnarray}
Thereby, we can determine whether a given function (A) is constant or
balanced with utilizing the state $|H\rangle_1|V\rangle_{1'}$.

%%%%%%%%%%%%%%%%%%%%%%%%%%%%%%%%%%%%%%%%%%%%%%%%%%%%%%%%
\subsection{Deutsch algorithm with a function B}

Similarly we can assume that the case where the state
$|V\rangle_1|H\rangle_{1'}$ is contributed to our Deutsch algorithm.
In this case, the unknown function (B) is determined in constant one
or balanced one. At the detector $D3$ the photon state becomes $
(|H\rangle_1 +|V\rangle_{1})|V\rangle_{D3}/\sqrt{2}. $ So a
vertically polarized photon is detected by the $D3$ detector. That
is,
\begin{equation}\sigma^{D3}_z=-1.  \label{Pauliz2}
\end{equation}

Therefore in the balanced function case, the changes of the
polarization states of the photons due to the dove prisms, HWP6 and
the last HWP($\theta=45^{\circ}$) are: $
(|V\rangle_{a}-|V\rangle_{b})/\sqrt{2}. $ Hence, we detect only
vertically polarized photons in $D4$. This implies
\begin{eqnarray}
(\sigma^{a}_z=\sigma^{b}_z=)\sigma^{D4}_z=-1.\label{PauliBz2}
\end{eqnarray}
Hence from Eqs.~(\ref{Pauliz2}) and (\ref{PauliBz2})
the value of observable $\sigma^{D3}_z \sigma^{D4}_z$ should be $+1$.
This is one of the result of a single run of the experiment, i.e.,
\begin{eqnarray}
\sigma^{D3}_z \sigma^{D4}_z=+1.\label{PaulifinalBz2}
\end{eqnarray}

In the constant function case, similarly we get after the dove
prisms, HWP6 and the last HWP($\theta=45^{\circ}$) $
(|H\rangle_{a}+|H\rangle_{b})/\sqrt{2}. $ And, we detect
horizontally polarized photons in $D4$. This implies
\begin{eqnarray}
(\sigma^{a}_z=\sigma^{b}_z=)\sigma^{D4}_z=+1.\label{PauliCz2}
\end{eqnarray}
Hence from Eqs.~(\ref{Pauliz2}) and (\ref{PauliCz2})
the value of observable $\sigma^{D3}_z\sigma^{D4}_z$ should be $-1$.
This is one of the success result of a single run of the experiment, i.e.,
\begin{eqnarray}
\sigma^{D3}_z\sigma^{D4}_z=-1.\label{PaulifinalCz2}
\end{eqnarray}
Thereby, we can determine whether a given function (B) is constant or
balanced with utilizing the state $|V\rangle_1|H\rangle_{1'}$.

Thus, we can determine whether either a given function (A) or (B) is
constant or balanced with utilizing EPR entanglement. Clearly, many
EPR experiments evaluate two functions simultaneously, i.e,
 Deutsch algorithm exhibiting a four to one speed up.
In the next
section, we assume the existence of experimental imperfections and
we present the method to determine the lower bound of the
success probability of our scheme presented by Fig.~\ref{setup3}.
Especially, a violation of Bell locality in the Hilbert space formalism
of quantum theory ensure the success probability is larger than $1/\sqrt{2}$.

%%%%%%%%%%%%%%%%%%%%%%%%%%%%%%%%%%%%%%%%%%%%%%%%%%%

\section{Bell operator analysis }\label{witness}
\begin{figure}
\begin{center}
\includegraphics[width=0.25\textwidth]{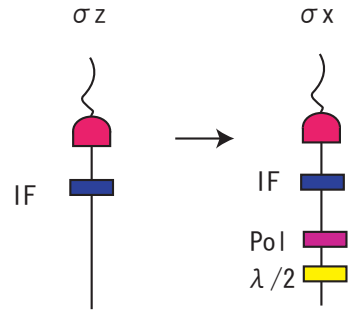}
\end{center}
\caption{A setup of measurement  of polarization in $z$ basis and in
$x$ basis} \label{sigmaX}
\end{figure}

In the previous section, we have assumed that the initial state is a
two-photon entangled state $ |\Psi\rangle=(|V\rangle_z |H\rangle_z
-|H\rangle_z|V\rangle_z)/\sqrt{2}$. We now insert a polarizer
oriented at $45^{\circ}$ and a HWP ($\lambda/2$ plate) in front of
each detector. See Fig.~\ref{sigmaX}. This allows the measurement of
polarized photon states described in polarized basis $x$. That is,
one can measure an observable $\sigma_x$ in this way. Due to the
feature of the initial state, the same situation occurs in the ideal
case. The situation is as follows. One can see
\begin{eqnarray}
|H\rangle_x=\frac{|H\rangle_z+|V\rangle_z}{\sqrt{2}},\nonumber\\
|V\rangle_x=\frac{|H\rangle_z-|V\rangle_z}{\sqrt{2}}.
\end{eqnarray}
Let us rewrite the initial state $|\Psi\rangle$ using $x$
polarization basis. We have $ |\Psi\rangle=(|V\rangle_x |H\rangle_x
-|H\rangle_x|V\rangle_x)/\sqrt{2}$. This implies that the scheme
mentioned in the preceding section works in the same way. However,
we have to take the imperfection of the photon detection and defects
in optical device into account.

Here we introduce Bell operators:
\begin{eqnarray}
&&B_A=\frac{1}{\sqrt{2}}(\sigma^{D1}_x\sigma^{D2}_x
+\sigma^{D1}_z\sigma^{D2}_z)\nonumber\\
&&B_B=\frac{1}{\sqrt{2}}(\sigma^{D3}_x\sigma^{D4}_x
+\sigma^{D3}_z\sigma^{D4}_z).
\end{eqnarray}
First of all, we check if both of the following Bell inequalities
\cite{BELL} are violated:
\begin{eqnarray}
|\langle B_A\rangle_{\rm avg}|\leq 1,~
|\langle B_B\rangle_{\rm avg}|\leq 1.
\end{eqnarray}
When both of the Bell inequalities are violated, we can ensure that
the success probability of  our Deutsch algorithm is larger than
$1/\sqrt{2}$ in the experiment as shown below.

We note here that if experimental error exits one could misjudge the
unknown function. For instance, it is possible that actually
observed data says that the unknown function is constant even though
the unknown function is in fact balanced. Such a wrong case occurs
when experimental error is larger than a half. Nevertheless, our
analysis rules out such a wrong case since a violation of two Bell
inequalities ensures the success probability of our scheme is larger
than $1/\sqrt{2}$.

In Table \ref{TABLE_ADV}, we summarize the relationship  between a
violation of Bell inequalities and the two types of functions ($A$)
and ($B$).

\begin{table}
\begin{tabular}{c | c c }
         &   $A$    &   $B$      \\
\hline
Balanced & $\langle B_A\rangle_{\rm avg}>1$
         & $\langle B_B\rangle_{\rm avg}>1$ \\
Constant & $\langle B_A\rangle_{\rm avg}<-1$
         & $\langle B_B\rangle_{\rm avg}<-1$
\end{tabular}
\caption{The rationship between the the violation  of Bell
inequalities and the two kinds of functions ($A$) and ($B$).}
\label{TABLE_ADV}
\end{table}

The situation is as follows.
%%%%%%%%%%%%%%%%%%%%%%%%%%%%%%%%%%%%%%%%%%%%%%%%%%%
First, we consider the case in which the unknown function (A) is
balanced. The fidelity to
$(|H\rangle_{D1}|H\rangle_{D2}+|V\rangle_{D1}|V\rangle_{D2})/\sqrt{2}$
in some quantum state $\rho$ (the success probability) is bounded as
\cite{NAGATA1}
\begin{eqnarray}
\langle B_A/\sqrt{2}\rangle \leq f^A_b \leq \frac{\langle
B_A/\sqrt{2}\rangle+1}{2}.
\end{eqnarray}
In an ideal case, we have $ \langle B_A\rangle=\sqrt{2}. $ In the
presence of experimental noise, we have
\begin{eqnarray}
\langle B_A \rangle=
\frac{1}{\sqrt{2}}(\langle\sigma_x^{D1} \sigma_x^{D2}\rangle_{\rm avg}+
\langle\sigma_z^{D1} \sigma_z^{D2}\rangle_{\rm avg}).
\end{eqnarray}
Hence, we can determine the range of the value of the success
probability $f^A_b$ in the presence of experimental noise. Thus, a
violation of Bell inequalities implies the success probability
$f^A_b$ is larger than $1/\sqrt{2}$ at least. We can analyze the
case where the unknown function (B) is balanced in a similar way.

%%%%%%%%%%%%%%%%%%%%%%%%%%%%%%%%%%%%%%%%%%%%%%%%%%%%%
Next, we consider the case where the unknown function (A) is
constant. The fidelity to $(|H\rangle_{D1}|V\rangle_{D2}
-|V\rangle_{D1}|H\rangle_{D2})/\sqrt{2}$ in some quantum state
$\rho$ (the success probability) is bounded as
\begin{eqnarray}
-\langle B_A/\sqrt{2}\rangle \leq f_c^A \leq \frac{-\langle
B_A/\sqrt{2}\rangle+1}{2}.
\end{eqnarray}
In ideal case, we have $ \langle B_A \rangle=-\sqrt{2}. $ In the
presence of the experimental noise, we have
\begin{eqnarray}
\langle B_A \rangle= -\frac{1}{\sqrt{2}}(\langle\sigma_x^{D1}
\sigma_x^{D2}\rangle_{\rm avg}+ \langle\sigma_z^{D1}
\sigma_z^{D2}\rangle_{\rm avg}).
\end{eqnarray}
Hence, we can determine the range of the value of the success
probability $f_c^A$ in the presence of experimental noise. Thus, a
violation of Bell inequalities implies the success probability
$f^A_c$ is larger than $1/\sqrt{2}$ at least. We can analyze the
case where the unknown function (B) is constant in a similar way.

We have two functions (A) and (B). The global success probability of
our Deutsch algorithm is given by
\begin{eqnarray}
P_{\rm success}=\frac{f^A_{k_1}+f^B_{k_2}}{2}~~(k_1,k_2\in\{b,c\}).
\end{eqnarray}
Clearly, this value $P_{\rm success}$ is equal to the global
probability with which a perfectly entangled state is detected.

As an example, suppose that the case where conditions $\langle
B_A\rangle_{\rm avg}>1$ and $\langle B_B\rangle_{\rm avg}<-1$ are
met. We can know that the unknown function (A) is balanced and (B)
is constant. In this case, our Deutsch scheme presented in
Fig.~\ref{setup3} succeeds with the probability of the value
$(\langle B_A/\sqrt{2}\rangle_{\rm avg}
+\langle-B_B/\sqrt{2}\rangle_{\rm avg})/2$ at least. This value is equal
to the lower bound of the probability with which a perfect entangled
state is detected, i.e., the lower bound of the global success
probability of our Deutsch algorithm. We can analyze other cases
(there are four cases in fact) in the same way. So we can
probabilistically determine whether each of two functions is
constant or balanced simultaneously, based on a violation of two
Bell inequalities and the evaluation of the success fidelity with
utilizing two-photon entangled state. This implies {\it
probabilistic} Deutsch algorithm exhibiting a four to one speed up
in ideal case. A violation of Bell locality in Hilbert space says
probabilistic Deutsch algorithm exhibiting a $2\sqrt{2}$ to one
speed up at least.

%%%%%%%%%%%%%%%%%%%%%%%%%%%%%%%%%%%%%%%%%%%%%%%%%%%%%%%%%%%%

\section{Summary and conclusion}\label{summary}
In summary, we have presented a linear-optical implementation of
quantum algorithm with the use of entanglement of photon states. For
the process of Deutsch algorithm, two-photon two-qubit entangled
states have been considered in conjunction with a polarization-based
C-NOT gate. The algorithm presented here is the only algorithm which
incorporates the Deutsch algorithm with a violation of Bell
inequalities, to date. A violation of Bell inequalities ensures the
success of probabilistic Deutsch algorithm with two unknown
functions which exhibits at least a $2\sqrt{2}$ to one speed-up
probabilistically. The global nonlocal effect leads us to make
quantum computer faster than usual ones.

\begin{acknowledgments}
This work has been supported by Frontier Basic Research Programs at
KAIST and K.N. is supported by the BK21 research professorship.
\end{acknowledgments}


\begin{thebibliography}{9}


\bibitem{NagataNakamura}
K. Nagata and T. Nakamura,
arXiv:0810.3134.

\bibitem{IONTRAPQC}
J. I. Cirac and P. Zoller, Phys. Rev. Lett. {\bf 74}, 4091 (1995).

\bibitem{ATOMQC}
D. Jaksch, Contemp. Phys. {\bf 45}, 367 (2004).

\bibitem{QuantumDotQC}
T. Hayashi, T. Fujisawa, H. D. Cheong, Y. H. Jeong, and Y. Hirayama,
Phys. Rev. Lett. {\bf 91}, 226804 (2003).

\bibitem{NMRQC}
N. A. Gershenfeld and I. L. Chuang,
Science {\bf 275}, 350 (1997).

\bibitem{SCQC}
Y. Nakamura, Yu. A. Pashkin, and J. S. Tsai,
Nature {\bf 398}, 786 (1999).

\bibitem{AhnScience2000}
J. Ahn, T. C. Weinacht, and P. H. Bucksbaum,
Science {\bf 287}, 463 (2000).

\bibitem{BhattacharyaPRL2002}
N. Bhattacharya, H. B. van Linden van den Heuvell, and R. J. Spreeuw,
Phys. Rev. Lett. {\bf 88}, 137901 (2002).

\bibitem{NC}
M. A. Nielsen and I. L. Chuang, {\it Quantum Computation and Quantum
Information} (Cambridge University Press, Cambridge, 2000).

\bibitem{Galindo}
A. Galind and M. A. Mart\'\i n-Delgado,
Rev. Mod. Phys. {\bf 74}, 347 (2002).

\bibitem{CRYPTOGRAPHY}
A. K. Ekert, Phys. Rev. Lett. {\bf 67}, 661 (1991);
V. Scarani and N. Gisin,
Phys. Rev. Lett. {\bf 87}, 117901 (2001);
A. Ac\'\i n, N. Gisin, and V. Scarani,
Quant. Inf. Comp. {\bf 3}, 563 (2003).

\bibitem{BRUKNER_PRL}
{\v C}. Brukner, M. \.Zukowski, J.-W. Pan, and A. Zeilinger,
Phys. Rev. Lett. {\bf 92}, 127901 (2004).


\bibitem{Oliveira}
A. N. de Oliveira, S. P. Walborn, and C. H. Monken,
J. Opt. B: Quantum Semiclass. Opt. {\bf 7}, 288-292 (2005).

\bibitem{Kim2003}
Y.-H. Kim,
Phys. Rev. A {\bf 67}, 040301(R) (2003).

\bibitem{Mohseni2003}
M. Mohseni, J. S. Lundeen, K. J. Resch, and A. M. Steinberg,
Phys. Rev. Lett. {\bf 91}, 187903 (2003).

\bibitem{BELL}
J. S. Bell, Physics (Long Island City, N.Y.) {\bf 1}, 195 (1964);
J. F. Clauser, M. A. Horne, A. Shimony, and R. A. Holt,
Phys. Rev. Lett. {\bf 23}, 880 (1969).


\bibitem{DEUTSCH}
D. Deutsch, {\it Proc. Roy. Soc. London Ser. A} {\bf 400}, 97
(1985).

\bibitem{NAGATA1}
K. Nagata, M. Koashi, and N. Imoto,
Phys. Rev. A {\bf 65}, 042314 (2002).




\bibitem{CNOT}
M. Fiorentino and F. N. C. Wong,
Phys. Rev. Lett. {\bf 93}, 070502 (2004).


\end{thebibliography}
\end{document}